\documentclass[a4paper,twocolumn,11pt,accepted=2024-03-04]{quantumarticle}
\pdfoutput=1
\usepackage{amsmath}
\usepackage{amssymb}
\usepackage[utf8]{inputenc}
\usepackage[colorlinks=true,citecolor=blue,linkcolor=blue,urlcolor=blue]{hyperref}
\usepackage{orcidlink}
\usepackage[numbers,sort&compress]{natbib}

\usepackage{verbatim}
\usepackage{float}
\usepackage{graphicx}
\usepackage{color}
\usepackage{physics}

\newcommand{\cop}{\hat{c}}
\newcommand{\cdop}{\hat{c}^\dag}
\newcommand{\Hop}{\hat{H}}

\newcommand{\mean}[1]{\left\langle{#1}\right\rangle}

\begin{document}

\title{Stabilization of Hubbard-Thouless pumps through nonlocal fermionic repulsion}
\author{Javier Arg\"uello-Luengo}
\affiliation{ICFO - Institut de Ciencies Fotoniques, The Barcelona Institute of Science and Technology, Av. Carl Friedrich Gauss 3, 08860 Castelldefels (Barcelona), Spain}
\orcid{0000-0001-5627-8907}
\author{Manfred\,J.\,Mark}
\affiliation{Institut f\"ur Quantenoptik und Quanteninformation, \"Osterreichische Akademie der Wissenschaften, Technikerstra{\ss}e 21a, 6020 Innsbruck, Austria}	
\affiliation{Institut f\"ur Experimentalphysik, Universit\"at Innsbruck, Technikerstra{\ss}e 25, 6020 Innsbruck, Austria}
\orcid{0000-0001-8157-4716}
\author{Francesca\,Ferlaino}
\affiliation{Institut f\"ur Quantenoptik und Quanteninformation, \"Osterreichische Akademie der Wissenschaften, Technikerstra{\ss}e 21a, 6020 Innsbruck, Austria}	
\affiliation{Institut f\"ur Experimentalphysik, Universit\"at Innsbruck, Technikerstra{\ss}e 25, 6020 Innsbruck, Austria}
\orcid{0000-0002-3020-6291}
\author{Maciej Lewenstein}
\affiliation{ICFO - Institut de Ciencies Fotoniques, The Barcelona Institute of Science and Technology, Av. Carl Friedrich Gauss 3, 08860 Castelldefels (Barcelona), Spain}
\affiliation{ICREA, Pg. Llu\'is Companys 23, 08010 Barcelona, Spain}
\orcid{0000-0002-0210-7800}
\author{Luca Barbiero}
\affiliation{Institute for Condensed Matter Physics and Complex Systems,
DISAT, Politecnico di Torino, I-10129 Torino, Italy}
\orcid{0000-0001-9023-5257}
\author{Sergi Juli\`a-Farr\'e}
\email{sergi.julia@icfo.eu}
\affiliation{ICFO - Institut de Ciencies Fotoniques, The Barcelona Institute of Science and Technology, Av. Carl Friedrich Gauss 3, 08860 Castelldefels (Barcelona), Spain}
\orcid{0000-0003-4034-5786}

\begin{abstract}
Thouless pumping represents a powerful concept to probe quantized topological invariants in quantum systems. We explore this mechanism in a generalized Rice-Mele Fermi-Hubbard model characterized by the presence of competing onsite and intersite interactions. Contrary to recent experimental and theoretical results, 
showing a breakdown of quantized pumping induced by the onsite repulsion, we prove that sufficiently large intersite interactions allow for an interaction-induced recovery of Thouless pumps.
 Our analysis further reveals that the occurrence of stable topological transport at large interactions is connected to the presence of a spontaneous bond-order-wave in the ground-state phase diagram of the model. Finally, we discuss a concrete experimental setup based on ultracold magnetic atoms in an optical lattice to realize the newly introduced Thouless pump.  Our results provide a new mechanism to stabilize Thouless pumps in interacting quantum systems.
\end{abstract}

\maketitle
\section{Introduction}
Since the discovery of the two-dimensional integer quantum Hall effect (IQHE)~\cite{Klitzing1980, TKNN1982}, the classification of phases of matter through global topological invariants has become an intense research area~\cite{RevModPhys.82.3045,chiu_2016}. Topological phases are of great interest both from a fundamental viewpoint, as they are not captured by the standard Ginzburg-Landau-Wilson theory~\cite{Landau_ssb,wilson_ssb} of spontaneously symmetry-broken phases and, potentially, at the technological level, as they represent promising tools for quantum metrology~\cite{Klitzing2017} and quantum computation~\cite{RevModPhys.80.1083}. While noninteracting topological matter can be accurately described in terms of the topological invariant and symmetries of single-particle bands, such classification becomes challenging in the presence of interactions~\cite{Rachel_2018}. 

In this context, Thouless pumps~\cite{Thouless1983_prb,QNiu_1984} have emerged as a powerful tool to characterize the topology of interacting one-dimensional systems through real-time dynamics, i.e., without relying on static groundstate properties. Specifically, these pumps represent a one-dimensional reduction of the IQHE, as they exploit the direct correspondence between quantized particle pumping and the topological invariant during cyclic adiabatic dynamics around a topological singularity. This scheme provides a natural theoretical and experimental framework to study the effect of interactions in topological systems undergoing such adiabatic transport dynamics. From a theoretical perspective, recent works showed that topological quantized transport can survive, and even be induced by many-body correlations in interacting bosonic systems~\cite{Berg2011,Greschner2020,hayward2018,mondal2021_inheritance, Lin2020_dimerized,Mondal2022,Kuno2020,padhan2023interacting}. In contrast, a
repulsive onsite Hubbard $U$ interaction~\cite{Nakagawa2018,Bertok2022} or electron-phonon coupling~\cite{Mondal2022_phononbreakdown,erratum_phononinduced} lead to a breakdown of the quantized Thouless pump in fermionic systems. At the experimental level, while initially conceived within the realm of solid-state systems~\cite{Thouless1983_prb,QNiu_1984}, the advent of analog quantum simulators~\cite{Feynman_1982,CiracZoller_reviewQS,Georgescu2014_reviewqs, Daley_practicalQS,Altman_reviewqs2021}, offering unprecedented levels of control and tunability, has revolutionized the study of Thouless pumps~\cite{Cooper_2019, AidelsburgerThouless2023,TopologicalPhotonicsreview_2019}. These advances have enabled the first experimental detection of topological quantized transport in photonic platforms~\cite{Kraus2012_pump,Cerjan2020Thouless,Jurguensen2021Thouless,TopologicalPhotonicsreview_2019,Jurgensen2023}, and in systems where ultracold bosonic~\cite{Lohse2016} and fermionic~\cite{Nakajima2016,Minguzzi2022,Walter2022,viebahn2023interactioninduced} atoms are trapped
in optical lattices~\cite{MaciejBook,Bloch_2008}.

A crucial advantage offered by ultracold atomic systems, in particular, is their impressive flexibility in engineering highly tunable interactions, as evidenced by the recent realization of longly sought interacting topological states of matter~\cite{Sompet2021,Léonard2023}. In the case of Thouless pumps, although the first experiments focused on single-particle transport~\cite{Lohse2016,Nakajima2016,Minguzzi2022}, the pioneering experiment~\cite{Walter2022}, consisting of a fermionic spin mixture of neutral atoms trapped in an optical lattice, reported the breakdown of Thouless pumps at large repulsive Hubbard $U$. This observation was in agreement with the mechanism described in Ref.~\cite{Nakagawa2018}, which explained this breakdown by the appearance of a gapless Mott insulator (MI) state, in which the transport of fermionic pairs is suppressed, and lead to the conclusion that for a dominant repulsive $U$ quantized charge pump is absent in the thermodynamic limit. Remarkably, the subsequent study~\cite{Bertok2022} 
 provided an alternative interpretation of this breakdown in terms of the splitting of the topological singularity,
 opening up the possibility to observe half-quantized transport by encircling the displaced singularities at large $U$ in small systems exhibiting a finite-size gap. 
 This latter exotic Thouless pump has been recently studied in the experiment~\cite{viebahn2023interactioninduced}, which showed that under realistic conditions half-quantized transport can be observed for single or few Thouless cycles, despite this effect being absent in the thermodynamic limit. While the abovementioned results allow for a deep understanding of the efficiency of Thouless pumps in systems characterized by local fermionic interactions, configurations where fermions interact through different and competing processes are unexplored and, therefore, novel phenomena might appear. 
 
 \begin{figure}[t!]
\includegraphics[width=0.9
\columnwidth]{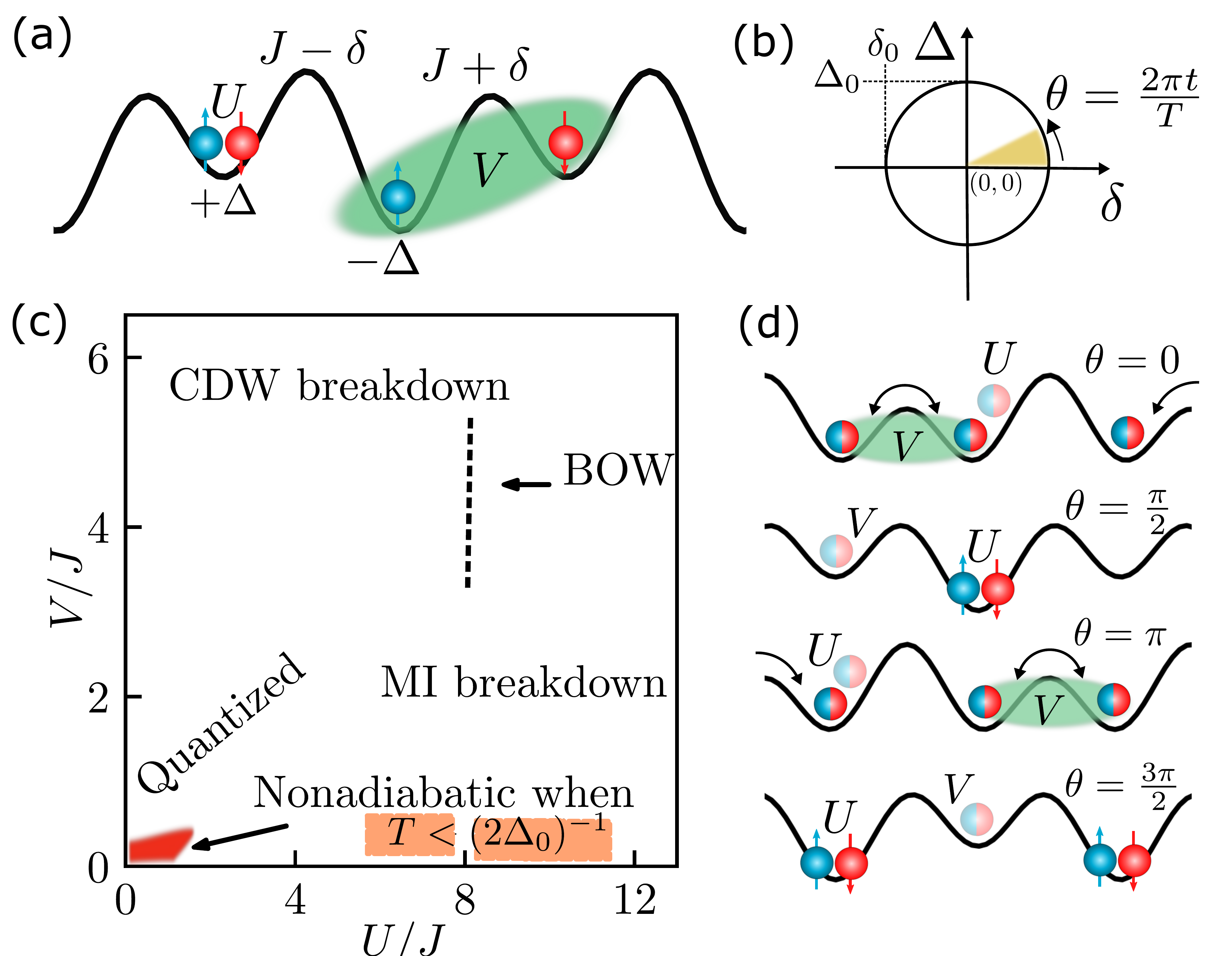}
\caption{(a) Rice-Mele extended Fermi-Hubbard model. (b) Thouless pump encircling the singularity $(\delta,\Delta)=(0,0)$. (c) Sketch of the adiabatic transport properties in the interacting $U-V$ plane, for fixed $\delta_0=0.9J,\ \Delta_0=J/2,\ TJ=50$. \textcolor{black}{For shorter times satisfying $T<(2\Delta_0)^{-1}$, Thouless pumping would become nonadiabatic in the noninteracting limit (red region). In the dark blue region, the ground state of the system is the BOW phase for $\Delta=\delta=0$~\cite{Ejima2007}.} (d) Expected phases during a Thouless pump. Note that $U$ ($V$) can prevent the interaction-induced CDW (MI) order that can break the pump at $\theta=0$ ($\theta=\pi/2$).}
\label{fig:intro}
\end{figure}
 In order to shed light on this regime, in this work we investigate fermionic Thouless pumps in the presence of both onsite and intersite
(nonlocal) interactions. This scenario results particularly important because, on the one hand, electrons in solid-state systems are subject to interactions with different range and, on the other, nonlocal fermionic interactions can be efficiently engineered in quantum simulators made of magnetic atoms~\cite{lahaye_physics_2009,Chomaz_2023}. To this end, we study Thouless pumps [Fig.~\ref{fig:intro}(b)] in a generalized Rice-Mele model with onsite $U$ and nearest-neighbor (NN) $V$ interactions, i.e., the Rice-Mele extended Fermi-Hubbard model sketched in Fig.~\ref{fig:intro}(a) using tensor network methods~\cite{schollwock_density-matrix_2011,tenpy} and exact diagonalization.  As main results [see Fig.~\ref{fig:intro}(c)], we find that (i) $V$ naturally allows one to preserve the adiabaticity of Thouless pumps in finite-time protocols, which is highly desirable from an experimental perspective, (ii) in contrast with previous experiments and numerical works~\cite{Walter2022,Nakagawa2018,Bertok2022}, we show that robust Thouless pumps can be realized even for large repulsive $U$, when realistic values of $V$ are considered. That is, we find a revival of Thouless pumps caused by the competition between $V$ and $U$, which is not limited to finite-size systems. And (iii), the quantization of Thouless pumps in this competition regime is connected to the presence of a largely debated spontaneously dimerized phase, the bond-order-wave (BOW)~\cite{Nakamura99, Nakamura2000,Jeckelmann2002,Sengupta2002,Sandvik2004,Zhang2004,Tam2006,Glocke2007,Ejima2007,Barbiero2014,julia-farre2021revealing}, in the $U-V$ groundstate phase diagram of the extended Fermi-Hubbard model. Finally, we also discuss a concrete experimental setup and protocol based on magnetic atoms in an optical lattice, to highlight that the newly introduced Thouless pumps in the regime of competing interactions can be observed in state-of-the-art dipolar quantum simulators.

This article is organized as follows. In Sections~\ref{sec:model} and ~\ref{sec:methods}, we describe, respectively, the Hamiltonian model and numerical methods used to analyze interacting Thouless pumps. In Section~\ref{sec:intersite}, we present the results of Thouless pumps in the presence of solely intersite interactions ($U=0$).  In Section~\ref{sec:competing}, we study the competition between intersite and onsite interactions. Section~\ref{sec:bow} contains the analysis of the connection between the quantization of Thouless pumps at moderately large interactions and the presence of a BOW phase in the phase diagram of a parent model. In Section~\ref{sec:experiment}, we discuss an experimental proposal with magnetic atoms, and we conclude in Section~\ref{sec:conclusions}.

\section{The Rice-Mele extended Fermi-Hubbard model}\label{sec:model}
We consider an extended version of the Rice-Mele Fermi-Hubbard model, sketched in Fig.~\ref{fig:intro}(a), describing a chain of length $L$ with $N$ spinful fermions, labeled by $\sigma=\uparrow,\downarrow$. Here we restrict to the half-filling case where both $N$ and the total magnetization $\hat{S}_z\equiv\sum_{i}(\hat{n}_{i,\uparrow}-\hat{n}_{i,\downarrow})/2$ are conserved with $N=L$ and $S_z=0$. The full Hamiltonian reads
\begin{equation}
\label{eq:Hamiltonian_DEFH}
\begin{split}
&\Hop=-\sum_{j=1, \sigma=\uparrow,\downarrow}^{L}\left[J-\delta (-1)^{j}\right](\cdop_{j,\sigma}\cop_{j+1,\sigma}+\textrm{H.c.}) +\\
&+\Delta\sum_j (-1)^j\hat{n}_j+U\sum_i\hat{n}_{j,\uparrow}\hat{n}_{j,\downarrow}
+V\sum_{j} \hat{n}_{j}\hat{n}_{j+1}.
\end{split}
\end{equation}
Here $J=1$ parametrizes the NN hopping with staggered dimerization strength $\delta$, and $\Delta$ is the strength of the staggered onsite energy. Concerning the interactions, $U>0$ accounts for an onsite repulsive Hubbard term, while $V>0$ describes the repulsion between fermions in NN sites. Note that $\Hop$ faithfully describes a fermionic dipolar mixture trapped in a one-dimensional optical lattice, in the presence of a superlattice giving rise to the staggered $\delta$ and $\Delta$ terms. 

The noninteracting limit of $\Hop$ corresponds to the Rice-Mele model~\cite{RiceMele1982}, or to the static Su-Schrieffer-Heegger (SSH) model~\cite{SSH1979} if chiral and inversion symmetries are further imposed, i.e., with $\Delta=0$. The SSH model exhibits symmetry-protected topological and trivial insulators for $\delta<0$ and $\delta>0$, respectively~\cite{Ryu2010_periodictable}. These phases are characterized by a global topological invariant, the winding number $\varphi$, which has an integer-quantized value $\varphi_{\text{topo}}=2$ ($\varphi_{\text{triv}}=0$) in the topological (trivial) sector~\cite{Manmana2012}. In the presence of inversion symmetry, $\varphi$ can only change at the singularity $\delta=0$, where a gap closing associated with a topological phase transition $|\Delta \varphi| =2$ takes place. 

The effect of interactions in the ground-state topological properties of $\Hop$ in the inversion-symmetric case ($\Delta=0$) has attracted great interest recently~\cite{Gurarie2011,Manmana2012,Yoshida2014,Wang2015,Ye2016,Sbierski2018,Barbiero2018,Le2020,Lin2020_interRM,Montorsi2022}. In a nutshell, the ground state of the system remains topologically protected at $\delta<0$ for sufficiently small values of $U$ and $V$, as these interactions preserve the protecting symmetries of the static SSH model. However, any finite $U$ ($V$) has an impact on the bulk-edge correspondence of the topological phase~\cite{Yoshida2014}, and only the edge modes in the charge (spin) sector preserve their topological degeneracy. Moreover, large interactions are in general detrimental to the topology. For sufficiently strong $V$, the system features a transition to a charge-density-wave (CDW), characterized by alternating empty and doubly occupied sites. Such a phase breaks the inversion and chiral symmetries that protect the topological phase, leading to a trivial band insulator. Concerning onsite interaction, for increasing $U$ the system evolves continuously towards a MI. While such a state does preserve the protecting symmetries, the bulk spin gap vanishes in the limit $U\to \infty$, and the degenerate spin edge states merge with the bulk modes.
Finally, notice that $\Hop$ can also exhibit a spontaneously dimerized insulator at $\delta=0$, the BOW phase, arising at the transition between competing orders. For instance, in the extended $J-U-V$ Fermi-Hubbard limit of $\Hop$, which is believed to capture the physics of several solid-state chains, a BOW appears in the phase diagram for moderate interactions~\cite{Nakamura99, Nakamura2000,Jeckelmann2002,Sengupta2002,Sandvik2004,Zhang2004,Tam2006,Glocke2007,Ejima2007,Barbiero2014,julia-farre2021revealing}, in a finite region between the CDW and the MI, preventing a direct phase transition between these two phases~\footnote{It is worth mentioning that another spontaneous BOW phase is also present in the ionic Hubbard limit of $\Hop$, namely at $\delta=V=0$, due to the competition between the onsite repulsion $U$ and the staggered potential $\Delta$, which we do not consider in this work. However, we note that this phase is not topologically protected, as the ionic Hubbard model breaks explicitly the protecting inversion and chiral symmetries. The latter explains the absence of quantized Thouless pumps in the regime of the spontaneous BOW phase of the ionic Hubbard model, described in Ref.~\cite{Nakagawa2018}, and its appearance in the regime of the chiral-symmetric spontaneous BOW phase of the extended Fermi-Hubbard model discussed here.}. Interestingly, such a BOW phase has the same properties as the trivial and topological ground states of the static SSH model, and therefore it can be viewed as an interaction-induced topological phase~\cite{julia-farre2021revealing}. \textcolor{black}{In this work, we show that the abovementioned richness of the ground state phase diagram of $\Hop$ in the SSH symmetric case ($\Delta=0$) also leads to a plethora of dynamical effects in the Rice-Mele regime ($\Delta \neq 0$), in which the adiabatic breaking of the protecting inversion symmetry can lead to quantized transport related to the groundstate topology. In particular, we study the resulting interacting Thouless pumps, with a focus on the unexplored effect of $V$ and its competition with the onsite term $U$.}
 
\section{Numerical methods for interacting Thouless pumps}\label{sec:methods}
To study Thouless pumps in systems described by $\Hop$, we consider the simultaneous periodic modulation in time of the bond superlattice $\delta$, and the staggered site potential $\Delta$, which breaks inversion and chiral symmetries. As sketched in Fig.~\ref{fig:intro}(b), this protocol allows for adiabatic transport by slowly encircling the singularity $\delta=0$.  We parametrize the Hamiltonian $\Hop(\theta)$ for a pump period as
\begin{equation}
\begin{split}
\delta (\theta) = \delta_0 \cos(\theta),\  \Delta (\theta) = \Delta_0 \sin(\theta),\ \theta & \in [0,2\pi].
\end{split}
\end{equation}
Here $\theta$ is related to the time $t$ by $\theta = 2\pi t/T$, where $T$ is the time of a single cycle, and we consider that at $t=0$ the system is in the ground state of $H(0)$. We are particularly interested in the charge that is pumped from one edge to the other of the chain during the cycle, which can be expressed as the integral over time of the instantaneous bulk current, that is:
\begin{equation}
\Delta Q_\theta = \int_0^{t(\theta)} \mathcal{J}(t')\, \text{dt}',
\end{equation}
where 
\begin{equation}
    \mathcal{J}(t)=\frac{i}{2}\sum_{j\in(0,1),\sigma} J_j\mean{\cdop_{L/2+j,\sigma}\cop_{L/2+j+1,\sigma}-\text{H.c.}},
\end{equation}
is the current operator averaged over a two-site unit cell in the bulk, and $J_j=(J\pm \delta)$. The connection between charge transport and the presence of a topological invariant comes from the fact that the transferred charge at the end of a cycle, $\Delta Q_{2\pi}$, can be explicitly written in terms of the change in the topological invariant across the topological singularity~\cite{thouless_quantized_1982}
\begin{equation}
    |\Delta Q_{2\pi}| = |\Delta \varphi|\in \mathbb{Z}.
    \label{eq:quantization_pump}
\end{equation}
We also define the regime of slow time dynamics through the condition $\left[T\min_\theta E_g(\theta)\right]^{-1}\ll 1$,
where $E_g(\theta)$ is the instantaneous bulk gap of $\Hop(\theta)$ during the pump. That is, the pump cycle frequency has to be much smaller than the lowest excitation of the system during the pump, to prevent excitations from higher bands. For instance, in the noninteracting Rice-Mele case, the adiabatic timescale is given by $\left[2T \min (\Delta_0,\delta_0)\right]^{-1}\ll 1$. In the more complex spinful interacting scenario, we consider the energy gaps at half filling ($N=L$) with respect to the ground state energy $E_0\equiv E_0(S_z=0)$. We define both the internal energy gaps $\Delta E_i$ in the unpolarized subspace, as well as the spin gaps  $\Delta E^s_i$ corresponding to flipping one spin as
\begin{equation}
\begin{split}
    \Delta E_i \equiv E_i(S_z=0)-E_0,\ \Delta E^s_i \equiv E_i(S_z=1)-E_0.
\end{split}
\end{equation}
To study the adiabatic dynamics generated by $\Hop(\theta)$, we fix the maximal bond dimerization to $\delta_0=0.9J$, which in the absence of interactions induces a large gap $E_g(0)=2\delta_0=1.8J$. We also consider relatively fast cycles of duration $TJ=50$, which are convenient for cold atom experiments with limited lifetimes. For the numerical calculations, we combine tensor network simulations in infinite systems using the infinite density-matrix-renormalization group (DMRG)~\cite{White1992,schollwock_density-matrix_2011} and time-evolving-block-decimation (TEBD) algorithms~\cite{itebd_vidal}, with a maximum bond dimension of $\chi_\text{max}=400$, and exact diagonalization in small periodic systems to estimate the bulk energy gaps. 

\section{Vanishing onsite interaction $U$}\label{sec:intersite}
\begin{figure}[t]
\includegraphics{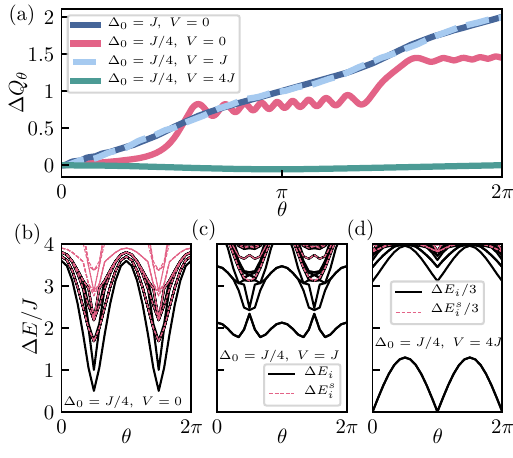}
\caption{ Effect of long-range interactions in Thouless pumps at $U=0$. (a) Evolution of the pumped charged for different values of $V$ and $\Delta_0$, simulated with the iTEBD algorithm. (b)-(c) First excitation energies of $\Hop(\theta)$ during the pumps with $\Delta_0=J/4$ for (b) $\ V=0$,  (c) $V=J$, and (d) $V=4J$, computed with exact diagonalization and $L=8$.}
\label{fig:results1}
\end{figure}

Let us first explore the effect of solely nonlocal $V$ interactions, i.e., at $U=0$, in a Thouless pump. As depicted in Fig.~\ref{fig:intro}(d), the main effect of a moderate $V$ term in our model is to suppress the uniform charge distribution in the regions $\theta=\pi/2$ and $\theta=3\pi/2$. This enhances the CDW order, and its associated insulating gap, in cooperation with the $\Delta$ term. A direct consequence is that finite values of $V$ can allow one to reduce the time period $T$ required for having adiabatic (and thus quantized) Thouless pumps. This behavior can be readily seen in Fig.~\ref{fig:results1}(a), which shows the evolution of the transferred charge $\Delta Q_\theta$ for different choices of $\Delta_0$ and $V$. We observe perfectly quantized pumps for the case $(\Delta_0,V)=(J,0)$, as expected for such an adiabatic noninteracting pump, and also for a finite interaction $(\Delta_0,V)=(J/4,J)$ (blue lines). In both cases, the final value of $\Delta Q_{2\pi}=2$ signals the presence of a topological singularity encircled by the closed adiabatic path, following Eq.~\eqref{eq:quantization_pump}. In contrast, note that, in the absence of interactions, the pump with $(\Delta_0,V)=(J/4,0)$ is not adiabatic and $\Delta Q_{2\pi}$ reaches a nonquantized value, as we observe strong oscillations of $\Delta Q_\theta$ in Fig.~\ref{fig:results1}(a) (pink line), caused by the coupling to excited states. In this regime of relatively small $\Delta_0=J/4$, we can understand the advantage of a moderate $V$ in preserving the adiabatic nature of the pump from the instantaneous gap behavior of $\Hop(\theta)$, shown in Fig.~\ref{fig:results1}(b)-(c). In the noninteracting case of Fig.~\ref{fig:results1}(b), the smallest gaps approach a zero value. In contrast, in Fig.~\ref{fig:results1}(c) one can see a clear enhancement of the gaps due to the finite $V$.

Importantly, note that while $V$ can be helpful to reduce the time cycles of the pump close to the noninteracting Rice-Mele limit, as sketched in the red-to-quantized transition of Fig.~\ref{fig:intro}(c), this is not true for arbitrarily large values of $V$. As sketched in Fig.~\ref{fig:intro}(c), at large $V$ the system exhibits a spontaneous CDW, and no transport is observed, as can be seen in Fig.~\ref{fig:results1}(a) for $V=4J$ (green line). In this case, the initially dimerized state at $\theta=0$, depicted in Fig.~\ref{fig:intro}(d), is replaced by the twofold degenerate ground state consisting of two states with the form $\ket{02\dots 02},\ket{20\dots 20}$, which have a gap from the rest of the spectrum during the pump, as shown in Fig.~\ref{fig:results1}(d). We note that, in fact, any finite CDW order in $\Hop(0)$ leads to the breakdown of the topological pump, as Eq.~\eqref{eq:quantization_pump} is only strictly valid when the closed cycle encounters inversion-symmetric points~\footnote{For the similar model considered in Ref.~\cite{Marks2021}, the authors  argued that, even in such a regime dominated by a local spontaneous charge order, one can observe quantized adiabatic pumps. We note that their scheme relies on the system being able to jump discontinuously from the low-entangled CDW ground states, with broken translational symmetry, to the highly entangled symmetric superposition of the two local CDW ground states. Therefore, while this can provide valuable insights of the underlying topology in numerical ground state simulations, such effects cannot be observed in real-time dynamics.}. \textcolor{black}{This can also be understood from the point of view of the noninteracting singularity, which becomes a twofold degenerate CDW line in the $\delta$ axis, and thus ceases being encircled by the closed adiabatic path, as depicted in Fig.~\ref{fig:singularities}(c)}.

Finally, we also note another limitation of the gap enhancement driven by the $V$ term when $\Delta_0\to 0$. One could think that, even for infinitesimally small $\Delta_0$, $V$ can amplify the CDW insulating order and drive an interaction-induced Thouless pump, as indeed suggested in Ref.~\cite{Marks2021}. However, we note that such an adiabatic scheme is not possible. In this limit, $\Hop$ preserves inversion symmetry and thus, it cannot induce net charge currents unless this symmetry is spontaneously broken in a thermodynamic phase transition. Such a transition to the spontaneous CDW implies an unavoidable gap closing leading to a dramatic breaking of the adiabatic condition.

\section{Interplay between onsite and nonlocal interactions}\label{sec:competing}
Let us now consider the combined effect of $U$ and $V$, which constitutes the primary motivation of this work, due to the possibility of simulating this regime of competing interactions with magnetic atoms in optical lattices. To this aim, here we focus on the regime of $U\gg\Delta_0$ by fixing $U=8J$ and $\Delta_0=J/2$. We note that, in the absence of $V$, this limit is already quite well understood~\cite{Nakagawa2018,Bertok2022,Walter2022}: $U$ brings the system into a MI which has a dramatic effect in Thouless pumps, as the suppression of double occupancies prevents the breaking of inversion symmetry generated by the $\Delta$ term during the pump at $\theta=\pi/2$ and $\theta=3\pi/2$ [see Figs.~\ref{fig:intro}(c)-(d)]. 
\begin{figure}[b!]
\includegraphics[width=\columnwidth]{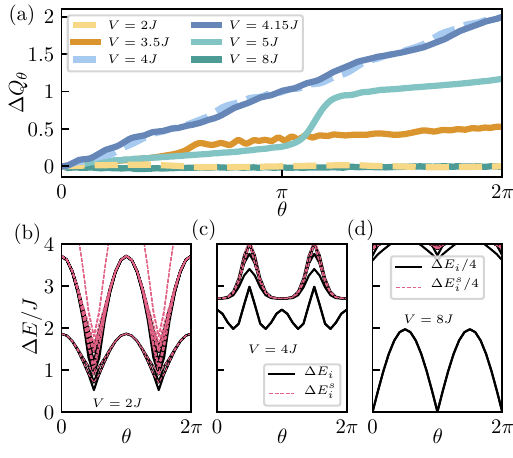}
\caption{Effect of intersite interactions in Thouless pumps at $U=8J$. Here we also fix $\Delta_0=J/2$. (a) Evolution of the pumped charged for different values of $V$, simulated with the iTEBD algorithm. (b)-(c) First excitation energies of $\Hop(\theta)$ during the pumps for (b) $\ V=2J$,  (c) $V=4J$, and (d) $V=8J$, computed with exact diagonalization and $L=8$.}
\label{fig:results2}
\end{figure}
Figure~\ref{fig:results2} proves how this scenario is modified when considering also the presence of $V$. In particular, Fig.~\ref{fig:results2}(a) and Figs.~\ref{fig:results2}(b)-(d) show, respectively, the pumped charge and instantaneous gaps for fixed $U=8J$ and $\Delta_0=J/2$ and different values of $V$. For a relatively small $V=2J$, we still observe in Fig.~\ref{fig:results2}(a) the absence of transport (yellow dashed line), showing that $U$ is still the dominant interaction in the system. The presence of a MI insulator is further confirmed by the strong reduction of the internal and spin gap observed in Fig.~\ref{fig:results2}(b).
However, when $V$ becomes larger, the transport is progressively enhanced and there is a revival of the quantized Thouless pump in a region around $V\sim U/2$ [blue lines of Fig.~\ref{fig:results2}(a)]. By observing Fig.~\ref{fig:intro}(d), we can explain such interaction-induced Thouless pumps from the simultaneous presence of large $U$ and $V$.  This leads to a competition between the breakdown caused by the MI~\cite{Nakagawa2018,Bertok2022,Walter2022}, and the CDW breakdown discussed in the previous section. That is, at such a large $U$, the $V$ term promotes double occupancies at $\theta=\pi/2$ and $3\pi/2$, and transport is recovered. At the same time, in this regime of strong NN repulsion, the $U$ term prevents any spontaneous CDW in the initial state at $\theta=0$. The absence of a MI or spontaneous CDW is also confirmed in Fig.~\ref{fig:results2}(c), which shows a robust gap during the pump. For an even larger $V\gtrsim 4.5J$, $U$ is not able to prevent the emergence of the spontaneous CDW order, and the quantized pump progressively disappears again, as shown in Fig.~\ref{fig:results2}(a). This effect is due to the appearance of a phase transition to the twofold degenerate CDW, as observed in Fig.~\ref{fig:results2}(d). Note that this CDW breakdown is analogous to the one occurring at $U=0$ described in Sec.~\ref{sec:intersite}, but here it takes place at a larger $V$ due to the competition induced by $U$. We also note that the behavior of Thouless pumps at $U=8J$ shown in Fig.~\ref{fig:results2} is found for other values of $U/J$ until a certain critical value $U_c\sim 10J$. Beyond this critical point, we find either a MI or a CDW breakdown of the pump, without an intermediate region of quantized transport. The latter is expected, as the tunneling dynamics do not play any role in the large interaction limit. This leads to a phase diagram for Thouless pumps in the $U/J-V/J$ plane of the Hamiltonian $\Hop$ in Eq.~\eqref{eq:Hamiltonian_DEFH}, for fixed values of $\delta_0/J=0.9$, $\Delta_0/J=0.5$, and $TJ=50$, sketched in Fig.~\ref{fig:intro}(c).
\begin{figure}[t!]
\includegraphics[width=\columnwidth]{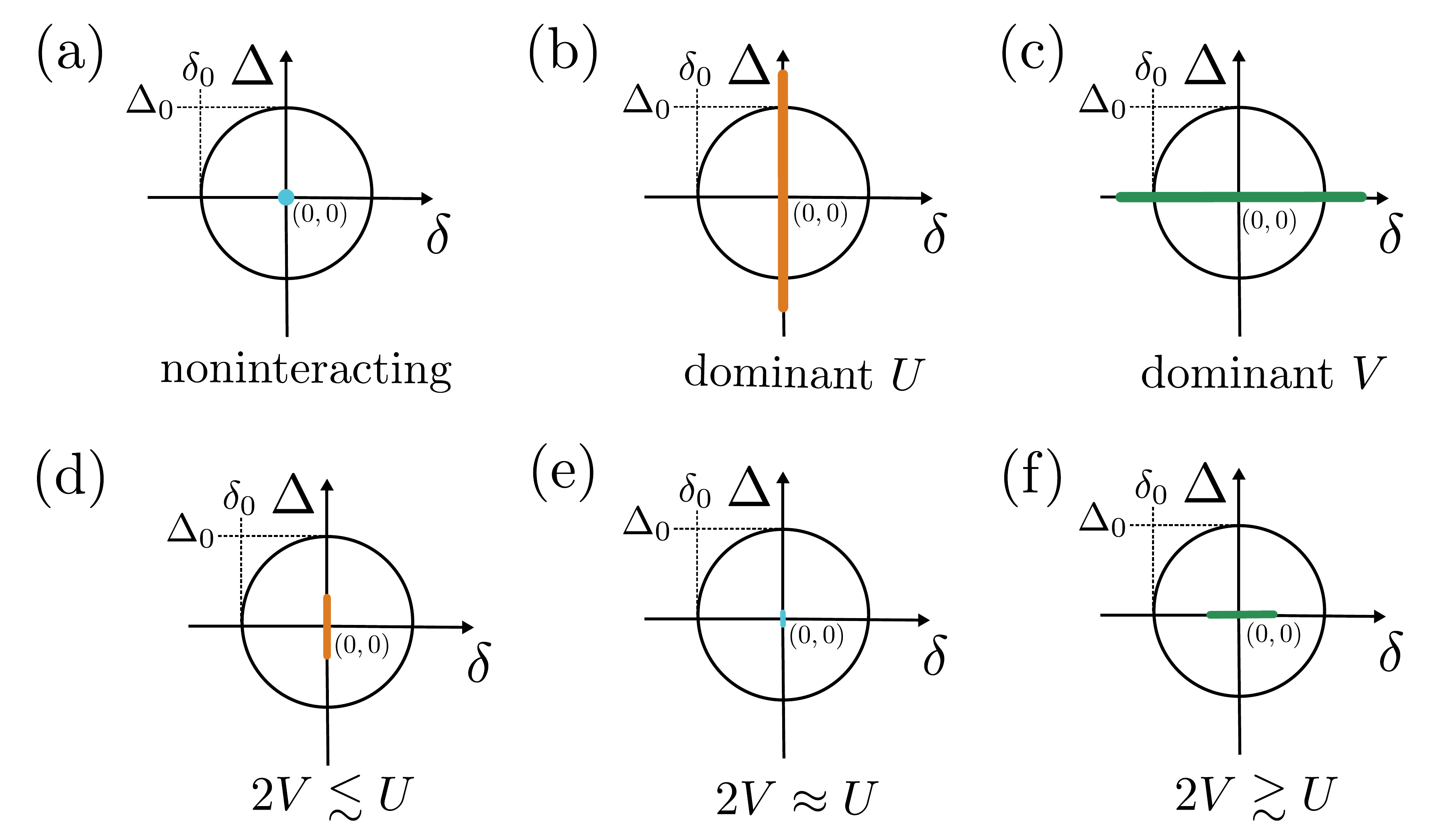}
\caption{Sketches of the ground-state singularities and gapless lines that appear in the $(\Delta,\delta)$ plane for different interaction regimes. The closed path used for Thouless pumps is also depicted. The orange lines in (b) and (d) represent a gapless MI. The green lines in (c) and (f) represent a twofold degenerate CDW. The blue short line in (e) corresponds to a twofold degenerate BOW. In panels (d)-(f) we considered the regime of finite hopping $J\sim U,V$.}
\label{fig:singularities}
\end{figure}

\textcolor{black}{Interestingly, one can also understand the interacting Thouless pumps shown in Fig.~\ref{fig:results2} and Fig.~\ref{fig:intro}(c) from the behavior of the topological singularity. To this end, in Fig.~\ref{fig:singularities} we sketch the evolution of the topological singularities in the $(\Delta,\delta)$ plane in the presence of interactions. For dominant $U$, the noninteracting singular point becomes the gapless MI vertical line~\cite{Bertok2022} of Fig.~\ref{fig:singularities}(b), which is not enclosed by the closed path. Alternatively, a dominant $V$ leads to the horizontal twofold degenerate line of Fig.~\ref{fig:singularities}(c), corresponding to the spontaneous CDW order. 
For a very large interaction scale $U,V \gg J$, increasing $V$ in the MI regime [Fig.~\ref{fig:singularities}(b)] leads to a first-order transition~\cite{Ejima2007} to the CDW regime of Fig.~\ref{fig:singularities}(c) around $U\approx 2V$. On the contrary, when interactions are of the order of the hopping, $U,V\approx J$, the enhanced quantum fluctuations lead to a competing regime in which such a first-order transition is replaced by a progressive reduction of the gapless lines, and an intermediate BOW phase~\cite{Ejima2007}, as sketched in Figs.~\ref{fig:singularities}(d)-(f). In this case, the closed path encircles the shorter gapless lines corresponding to the MI, BOW or CDW regimes, and quantized transport is recovered.}  

Remarkably, the above discussion shows that the behavior of Thouless pumps in the $U/J-V/J$ phase diagram sketched in Fig.~\ref{fig:intro}(c) is very reminiscent of the ground-state phase diagram of the extended Fermi-Hubbard ($J-U-V$) model~\cite{Nakamura99, Nakamura2000,Jeckelmann2002,Sengupta2002,Sandvik2004,Zhang2004,Tam2006,Glocke2007,Ejima2007,Barbiero2014,julia-farre2021revealing}, where the spontaneous BOW phase appears in a similar region as the quantized Thouless pumps. In what follows, we explore this connection and prospects to use it as a tool for detecting the spontaneous BOW phase.
\section{Thouless pumps and the bond-order-wave phase}\label{sec:bow} 
The fact that, in the regime given by $V\sim U/2,\ U<10J$, we can recover quantized Thouless pumps by increasing the strength of the $V$ interaction at fixed $U$ cannot be understood from a simple noninteracting picture. However, we can explain this phenomenon by noting that, in the absence of the $\Delta,\delta$ terms used during the pump, the ground state of the Hamiltonian~\eqref{eq:Hamiltonian_DEFH} is a spontaneous BOW phase~\cite{Nakamura99, Nakamura2000,Jeckelmann2002,Sengupta2002,Sandvik2004,Zhang2004,Tam2006,Glocke2007,Ejima2007,Barbiero2014} also in the regime $V\sim U/2,\ U<10J$, as sketched in Fig~\ref{fig:intro}(c). In such a phase, which is also induced by the competition between the MI and CDW orders, there is a ground-state bulk degeneracy between a topological and a trivial state~\cite{julia-farre2021revealing}. From the perspective of Thouless pumps, in the BOW region, the main effect of $\delta$ is to break the degeneracy between the two spontaneous BOW ground states and to control/guide its spontaneous symmetry breaking during the pump, such that the system is able to explore the trivial and topological dimerized sectors. 
\begin{figure}[b!]
\includegraphics[width=\columnwidth]{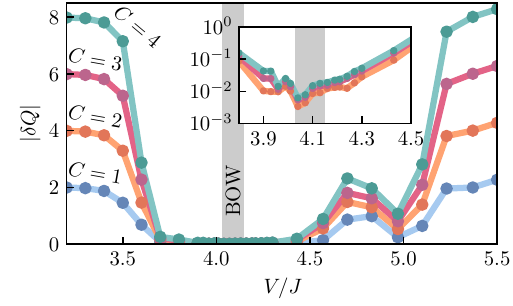}
\caption{(a) Accuracy of Thouless pumps (iTEBD) as a function of the nonlocal interaction $V/J$ and at different cycle numbers $C$, for fixed $U=8J$ and $\Delta_0=J/2$. We only observe quantized pumps independent of $C$ around the narrow region where the groundstate of the $J-U-V$ model \textcolor{black}{(with $\delta=\Delta=0$)} is the spontaneous BOW phase (gray shaded area, extracted from Ref.~\cite{Ejima2007}). \textcolor{black}{The latter can be more clearly seen with the inset logarithmic scale. The nonomonotonic behavior of $|\delta Q|$ in the window $V/J \in (4.5, 5.5)$ is related to the spontaneous CDW phase transition, as detailed in the main text.}}
\label{fig:detection_bow}
\end{figure}
Interestingly, one can therefore interpret the quantized Thouless pumping as an indirect proof of the presence of a BOW phase, whose detection and engineering remain elusive in cold atoms experiments, despite recent theoretical proposals~\cite{Loida2017}.

To illustrate such a relation more explicitly, in Fig.~\ref{fig:detection_bow} we plot the breaking of the quantization in Thouless pumps consisting of $C$ consecutive periodic cycles, defined as $\delta Q\equiv |\Delta Q_{2\pi C}|-2C$, for a fixed $U=8J$ and as a function of $V$, which corresponds to the dashed line in Fig.~\ref{fig:intro}(c). Remarkably, we observe perfectly quantized and cycle-independent transport, i.e.,  $|\delta Q|=0$, for Thouless pumps in the region $V/J\in(3.9,4.3)$ around the narrow window of the BOW phase $V/J\in(4,4.15)$. As sketched in Fig.~\ref{fig:intro}(c), the same behavior is also observed for other values of $U/J$.  \textcolor{black}{Two important comments are in order at this point, concerning the quantitative behavior of the curves in Fig.~\ref{fig:detection_bow}. In first place, note that the fact that the BOW region does not coincide exactly with the region of quantized pumps is expected. This is because, in the adiabatic dynamics considered here, the staggered field $\delta$ ($\Delta$) used to perform the pump prevents the CDW (MI) 
in a slightly wider range of $V$ values than in the ground-phase diagram of the model in the absence of such guiding fields~\cite{Ejima2007}. In second place, the nonmonotonic behavior of $|\delta Q|$ after the quantized plateau in Fig.~\ref{fig:detection_bow} is explained by the fact that there is a phase transition to a local CDW order at $V/J\approx 4.75$. At this critical point, $|\delta Q|$ exhibits a local maximum due to the strong deviation from the adiabatic regime in our finite-time protocol. The latter also leads to dynamical results that slighlty depend on the bond dimension $\chi_\text{max}$. For $V$ larger than the critical point, the system reenters the adiabatic regime due to the insulating nature of the CDW phase. Inside this CDW phase, increasing $V$ enhances the gap by moving away from the critical point, thus favoring adiabatic transport, but it also enhances the amplitude of the CDW spontaneous order. While the competition between these two phenomena initially leads to the counterintuitive  
decrease of $|\delta Q|$ for increasing $V$, 
we emphasize that in this CDW regime  Thouless pumps do not encircle a topological singularity and cannot be associated to any topological invariant. That is, $|\delta Q|$ is cycle-dependent and not quantized. Further increasing $V/J$ progressively leads to a maximally polarized CDW phase, suppressing any hopping (transport) process during the pump. }

As a general remark, note that the scheme depicted in Fig.~\ref{fig:detection_bow} overcomes the main problems that make it difficult to detect the BOW phase of the extended Fermi-Hubbard model from direct measurements. First, with Thouless pumps the topology of the interacting model~\eqref{eq:Hamiltonian_DEFH} can be probed by measuring the average global charge transport in the system, a topological property robust to small imperfections such as local disorder or low finite temperature. In contrast, a direct measurement scheme of the topological nature of the BOW phase would require, e.g., measuring nonlocal correlators~\cite{Barbiero2013,julia-farre2021revealing} with single-site resolution through a quantum gas microscope~\cite{Bakr2009,Endres2011,Hilker484}. Similarly, even a direct measurement of the BOW local order parameter requires involved probes capable of measuring the bond density in-situ, or superlattice modulation spectroscopy~\cite{Loida2017}. Second, the spontaneous nature of the BOW phase, and its very small intrinsic spin gap, require very small temperatures and complex state-preparation protocols to avoid the presence of solitonic excitations. On the contrary, in Thouless pumps the spin gap is enhanced due to the presence of the bond guiding field $\delta$, which also lifts the groundstate degeneracy of the BOW phase, thus preventing the appearance of solitons. Third, as shown in Fig.~\ref{fig:intro}(c) and Fig.~\ref{fig:detection_bow}, within experimental errors quantized pumps are observed in a much wider parameter region than the BOW phase. 
\section{Experimental proposal with magnetic atoms}\label{sec:experiment}
A prime candidate for an experimental realization of the Rice-Mele extended Fermi-Hubbard Hamiltonian~\eqref{eq:Hamiltonian_DEFH} are quantum simulators using ultracold lanthanide atoms in optical lattices~\cite{Patscheider2020,su2023dipolar}. For erbium, the preparation of a two-component Fermi gas in a lattice with long lifetimes and broad interspin Feshbach resonances for interaction control has been already demonstrated~\cite{Baier2018}. To manipulate the control parameters $\Delta$ and $\delta$ of the Thouless pump, a superlattice similar to experiments with neutral atoms can be used, see e.g., Refs.~\cite{Nakajima2016,Walter2022}. Such a superlattice can also be used to prepare the initial state e.g., a dimerized chain along the superlattice direction. As the proposed scenario is based on single one-dimensional systems, experimental setups need sufficient spacing between individual tubes to avoid strong coupling between the tubes due to the long-range character of the dipole-dipole interaction. This can be realized either by choosing a larger lattice spacing perpendicular to the superlattice, or by isolating single stripes in a two-dimensional lattice plane via the removal of atoms in between using single-site addressing of a quantum gas microscope.

 As already discussed in Ref.~\cite{julia-farre2021revealing}, taking realistic values currently achievable in experiments should allow one to reach the quantized pump regime with e.g.,  $U=4J$ and $V=2J$ [see Fig.~\ref{fig:intro}(c)] using erbium in a lattice with $532\,$nm unit cell ($266\,$nm short lattice spacing), giving a maximum interaction strength of $V/h\approx30\,$Hz. This can be further increased by using dysprosium and shorter wavelength lattices as planned for future experiments~\cite{Fraxanet2021,sohmen2023shipinabottle}, reaching interaction strengths of up to $V/h\approx200\,$Hz. To be able to observe the revival it would be preferential to probe the dynamics with and without $V$. To probe the dynamics without $V$, one can either rotate the quantization axis to the magic angle, fully eliminating $V$, 
 or scale $U$ and $J$ up considerably $(V\ll J,U$) to minimize the influence. \textcolor{black}{The regime $V\gg U$ can also be achieved, since $U$ can be tuned independently by means of Feshbach resonances~\cite{PhysRevA.79.013622}}. The induced transport due to the Thouless pump mechanism can be directly measured using the center-of-mass displacement of the atoms using either high-resolution in-situ imaging (see e.g.~\cite{Nakajima2016}) or a quantum gas microscope. For $J=2\pi\times 100\text{Hz}$, the adiabatic condition $TJ=50$ used in our numerical simulations would be achieved for realistic pump periods of half a second. We also expect that the averaged center-of-mass displacement is robust against defects appearing during the dynamics, making the lifetime requirements less stringent.
\section{Conclusions and outlook}\label{sec:conclusions}
We investigated the appearance of quantized topological transport, i.e., Thouless pumping, in a Rice-Mele extended Fermi-Hubbard model where fermions are subject to competing interactions of different ranges. Contrary to previous analysis showing a detrimental effect of onsite repulsive interactions to quantized Thouless pumps~\cite{Nakagawa2018,Bertok2022,Walter2022}, we unveiled that including nonlocal repulsions leads to novel phenomena.
 
 Specifically, in configurations with only intersite repulsive interactions $V$, we showed that a moderate fermionic repulsion favors the presence of quantized Thouless pumps. We proved this phenomenon to be totally induced by this specific interaction which tends to increase the instantaneous energy gaps and thus stabilize the system topology. In contrast, at large nonlocal interactions, we find an expected breakdown of the transport due to the appearance of a charge-density-wave. 
 
 When further considering a repulsive onsite Hubbard $U$, which is known to destroy the topological transport~\cite{Nakagawa2018,Bertok2022,Walter2022}, we still identified a sizeable region where topological transport persists. 
Here the nonlocal repulsion counteracts the destructive role of a moderately large local repulsion and vice versa, and therefore the quantized transport remains stable. Noticeably, this competing mechanism allows for a robust quantized transport at interactions up to one order of magnitude larger than the noninteracting energy, i.e., at $U/J\sim 10$.
 
 Interestingly, this stabilization of topological Thouless pumps through competing interactions
 is reminiscent of the ground state properties of the parent extended Fermi-Hubbard model. There, the presence of moderate and competing contact and intersite interactions favors the appearance of a spontaneous bond-order-wave phase, whose topological properties have been recently revealed \cite{julia-farre2021revealing}. Based on this fact, we thus expect that quantized topological transport can take place in different strongly correlated models where competing interactions give rise to spontaneously generated bond-order-wave regimes~\cite{Mondal2022}. Beyond our numerical analysis, we have been further able to propose a realistic experimental scheme where our findings can be probed. The latter relies on trapping an ultracold mixture of magnetic atoms in an optical lattice. The advantage of using such particles comes from the fact that they allow for a fine-tuning of both contact and intersite interaction,  thus providing access to simulate the Hamiltonian of Eq.~\eqref{eq:Hamiltonian_DEFH} in a wide regime of parameters. Importantly, note that these configurations where fermions are subject to local and nonlocal repulsions are the most common in solid-state platforms. Moreover, whilst we focused on repulsive interactions and charge transport, our numerical analysis and experimental proposal could be adapted to study the role of attractive terms and spin transport in the same model. We, therefore, believe that our results can have an important impact on a large variety of physical systems, thus representing an essential step towards a complete understanding of many-body phases of matter characterized by nontrivial topological properties.

\textbf{Acknowledgments}.
The authors acknowledge fruitful discussions with T. Esslinger, Z. Zhu, A. Dauphin, F. Heidrich-Meisner, A. A. Aligia, and E. Bertok. The DMRG and TEBD calculations were performed using the TeNPy Library~\cite{tenpy}. ICFO group acknowledges support from: ERC AdG NOQIA; MICIN/AEI (PGC2018-0910.13039/501100011033, CEX2019-000910-S/10.13039/501100011033, Plan National FIDEUA PID2019-106901GB-I00, FPI; MICIIN with funding from European Union NextGenerationEU (PRTR-C17.I1): QUANTERA MAQS PCI2019-111828-2); MCIN/AEI/ 10.13039/501100011033 and by the “European Union NextGeneration EU/PRTR" QUANTERA DYNAMITE PCI2022-132919 within the QuantERA II Programme that has received funding from the European Union’s Horizon 2020 research and innovation programme under Grant Agreement No 101017733Proyectos de I+D+I “Retos Colaboraci\'on” QUSPIN RTC2019-007196-7); Fundaci\'o Cellex; Fundaci\'o Mir-Puig; Generalitat de Catalunya (European Social Fund FEDER and CERCA program, AGAUR Grant No. 2021 SGR 01452, QuantumCAT \ U16-011424, co-funded by ERDF Operational Program of Catalonia 2014-2020); Barcelona Supercomputing Center MareNostrum (FI-2023-1-0013); EU (PASQuanS2.1, 101113690); EU Horizon 2020 FET-OPEN OPTOlogic (Grant No 899794); EU Horizon Europe Program (Grant Agreement 101080086 — NeQST), National Science Centre, Poland (Symfonia Grant No. 2016/20/W/ST4/00314); ICFO Internal “QuantumGaudi” project; European Union’s Horizon 2020 research and innovation program under the Marie-Sk\l odowska-Curie grant agreement No 101029393 (STREDCH) and No 847648 (“La Caixa” Junior Leaders fellowships ID100010434: LCF/BQ/PI19/11690013, LCF/BQ/PI20/11760031,LCF/BQ/PR20/
11770012, LCF/BQ/PR21/11840013). Views and opinions expressed are, however, those of the author(s) only and do not necessarily reflect those of the European Union, European Commission, European Climate, Infrastructure and Environment Executive Agency (CINEA), nor any other granting authority. Neither the European Union nor any granting authority can be held responsible for them. The Innsbruck group acknowledges support from a QuantERA grant MAQS from the Austrian Science Fund (FWF No. I4391-N). L. B. acknowledges financial support within the DiQut Grant No. 2022523NA7  funded by European Union – Next Generation EU,  PRIN 2022 program (D.D. 104 - 02/02/2022 Ministero dell’Università e della Ricerca).
\bibliographystyle{apsrev4-1}

\end{document}